\documentclass[12pt]{article}

\usepackage{amssymb}

\begin{document}

\title{Edge currents in the absence of edges}
\author{P.~Exner$^{a,b}$, A.~Joye$^{c}$, and H.~Kova\v{r}\'{\i}k$^{a,d}$,}
\date{}
\maketitle

\begin{flushleft}
 {\em a) Department of Theoretical Physics, Nuclear Physics Institute,
 Academy \phantom{a) }of Sciences, 25068 \v{R}e\v{z} near Prague
 \\ b) Doppler Institute, Czech Technical University, B\v{r}ehov\'{a}
 7, 11519 Prague, \phantom{a) }Czech Republic
 \\ c) Institut Fourier, Universit\'{e} de Grenoble 1, 38402 Saint-Martin
 d'Heres, \phantom{a) }France \\
 d) Faculty of Mathematics and Physics, Charles University,
 \phantom{a) }V~Hole\v{s}ovi\v{c}k\'ach~2, 18000 Prague
 \\ \quad{\em exner@ujf.cas.cz, joye@ujf-grenoble.fr,
 kovarik@ujf.cas.cz}}
\end{flushleft}

\begin{abstract}
We investigate a charged two-dimensional particle in a homogeneous
magnetic field interacting with a periodic array of point
obstacles. We show that while Landau levels remain to be
infinitely degenerate eigenvalues, between them the system has
bands of absolutely continuous spectrum and exhibits thus a
transport along the array. We also compute the band functions and
the corresponding probability current.
\end{abstract}

\vspace{8mm}

\noindent The fact that the presence of boundaries can induce a
transport in a system with a homogeneous magnetic field is known
for long \cite{hal,dos} and has numerous consequences in solid
state physics. The mentioned pioneering papers were followed by
tremendous number of studies in which the magnetic transport was
analyzed numerically in various models as well as experimentally.
The obstacles at which the particle ``bounces" need not be hard
walls but also objects with openings such as various antidot
lattices; for a sample of literature see, e.g.,
Refs.~\cite{fgk,zmh}, the following papers by the same authors,
and references therein. The closest to the subject of this letter
is a very recent paper \cite{ue} where the array of obstacles is
strictly one-dimensional and there are no boundaries to help the
transport.

One can say therefore that at the heuristic level most aspects of
the two-dimensional magnetic transport are understood. With this
fact and the extensive literature in mind it is a bit surprising
how little attention was paid during a decade and a half to a
strict derivation of transport properties from the first
principles -- which is after all the {\em raison d'}$\hat{e}${\em
tre} of theoretical physics. The situation concerning this aspect
of the problem changed recently where several rigorous studies
appeared. It was shown, e.g., that the edge currents in a
halfplane survive a mild disorder so that away of the Landau
levels the spectrum remains absolutely continuous \cite{bip,fgw1},
and that the result extends to planar domains containing an open
wedge \cite{fgw2}.

Recall that the wall producing the edge states need not be of a
potential type. It is known, e.g., that a step of the magnetic
field or another variation exhibiting a translational symmetry
will smear again the Landau levels into a continuous spectrum
\cite{iwa,map}. Similarly to the usual edge states, this type of
propagation allows for a classical explanation in terms of the
cyclotronic radius changing with the magnetic field -- see
Ref.~\cite{cycon}, Sec.~6.5.

The aim of the present letter is to contribute to this series of
rigorous studies with a simple solvable model in which a charged
quantum particle in the plane exposed to a homogeneous magnetic
field of intensity $B$ perpendicular to the plane interacts with a
periodic array of point obstacles situated at the $x$ axis and
described by $\delta$ potentials. We shall show that the model
exhibits a magnetic transport of which we can with a license say
that it is a purely quantum effect in the sense that a quantum
particle propagates despite the fact that its classical
counterpart moves on localized circular trajectories -- apart of a
zero-measure family of the initial conditions. The delta
potentials are certainly an idealization; a more realistic model
with potential-type obstacles will be discussed in a forthcoming
paper. Let us mention, however, that point interactions play a
distinguished role being the only obstacles that can preserve the
Landau levels in the spectrum.

Using the Landau gauge, we can write the Hamiltonian formally as
\begin{equation} \label{Hamiltonian}
H_{\alpha,\ell} = (-i\partial_x +By)^2 -\partial_y^2 +\sum_j
\tilde\alpha \delta(x\!-\!x_0\!-\!j\ell)\,,
\end{equation}
where $\ell>0$ is the array spacing. Since we are interested
mainly in the essence of the effect, we use everywhere
rationalized units $\hbar =c =e =2m =1$. The interaction term, in
particular the formal coupling constant $\tilde\alpha$, needs an
explanation, since the two-dimensional $\delta$ potential is an
involved object. We adopt here the conventional definition from
Ref.~\cite{aghh} which determines the latter by means of the
boundary conditions
\begin{equation} \label{bc}
L_1(\psi,\vec a_j)+ 2\pi\alpha L_0(\psi,\vec a_j)=0\,, \quad
j=0,\pm1,\pm2,\dots
\end{equation}
with $\vec a_j:= (x_0\!+\!j\ell,0)$, where $L_k$ are the
generalized boundary values
\begin{equation}
L_0(\psi,\vec a) := \lim_{|\vec x-\vec a|\to 0}\, {\psi(\vec
x)\over \ln |\vec x\!-\!\vec a|}\,, \; L_1(\psi,\vec a) :=
\lim_{|\vec x-\vec a|\to 0} \Bigl\lbrack \psi(\vec x)\!-\!
L_0(\psi,\vec a)\, \ln |\vec x\!-\!\vec a| \Bigr\rbrack\,,
\end{equation}
and $\alpha$ is the (rescaled) coupling constant; the free
(Landau) Hamiltonian corresponds to $\alpha=\infty$. Recall that
since the magnetic field amounts locally to a regular potential in
the s-wave subspace, the non-magnetic boundary conditions of
Ref.~\cite{aghh} need not be modified -- see, e.g.,
Ref.~\cite{bc}.

The difference between $\alpha$ and $\tilde\alpha$ reflects the
nontrivial way in which the two-dimensional point interaction
arises in the  limit of scaled potentials. Due to the coupling
constant renormalization, a caution is required when interpreting
spectral properties of such a Hamiltonian. On the other hand, the
two-dimensional point interaction yields a {\em generically}
correct description of low-energy scattering which can be tested
in experiments -- see, e.g., a fresh example in \cite{es}. To
understand the above remark about the purely quantum nature of the
transport here, recall the well-known expression for the
scattering length which shows that the obstacles have a ``point"
character if $e^{2\pi\alpha}\ll\ell$, i.e., if the point
interaction is strong enough. Below we shall show that the
transport exists here for {\em any} finite value of $\alpha$.

Using the periodicity, we can perform the Bloch decomposition in
the $x$ direction writing
\begin{equation}
H_{\alpha,\ell} = {\ell\over 2\pi}\,
\int_{|\theta\ell|\le\pi}^{\oplus} H_{\alpha,\ell}(\theta)
\,d\theta\,,
\end{equation}
where the fiber operator $H_{\alpha,\ell}(\theta)$ is of the form
(\ref{Hamiltonian}) on the strip $0\le x\le\ell$ with the boundary
conditions
\begin{equation} \label{Bloch bc}
\partial_x^i\psi(\ell-,y) = e^{i\theta\ell} \partial_x^i\psi(0+,y)\,,
\quad i=0,1\,.
\end{equation}
The Green's function of the operator $H_{\alpha,\ell}(\theta)$ is
given by means of the Krein formula \cite[App.A]{aghh},
\begin{eqnarray} \label{Krein}
\lefteqn{(H_{\alpha,\ell}(\theta)\!-\!z)^{-1}(\vec x,\vec
x^{\prime}) = G_0(\vec x,\vec x^{\prime};\theta,z)} \nonumber
\\&&  +(\alpha\!-\!\xi(\vec a_0;\theta,z) )^{-1} G_0(\vec x,\vec
a_0;\theta,z) G_0(\vec a_0,\vec x^{\prime};\theta,z)\,,
\end{eqnarray}
where $G_0$ is the free Green's function and
\begin{equation} \label{xi}
\xi(\vec a;\theta,z) := \lim_{|\vec x-\vec a|\to 0}\,
\left(G_0(\vec a,\vec x;\theta,z)- {1\over 2\pi}\,\ln|\vec
x\!-\!\vec a| \right)
\end{equation}
is its regularized value at the point $\vec a$. The Bloch
conditions (\ref{Bloch bc}) determine eigenvalues and
eigenfunctions of the transverse part of the free operator,
\begin{equation} \label{Bloch ev}
\mu_m(\theta) = \left({2\pi m\over\ell} + \theta \right)^2\!, \;
\eta_m^{\theta}(x) = {1\over\sqrt{\ell}}\, {\rm e}^{i(2\pi
m+\theta\ell)x/\ell}\,,
\end{equation}
where $m$ runs through integers. Then we have
\begin{equation}
G_0(\vec x,\vec x^{\prime};\theta,z) = - \sum_{m=-\infty}^{\infty}
{u_m^{\theta}(y_<) v_m^{\theta}(y_>) \over W(u_m^{\theta},
v_m^{\theta})}\, \eta_m^{\theta}(x)
\overline{\eta_m^{\theta}}(x^{\prime})\,,
\end{equation}
where $y_<, y_>$ is the smaller and larger value, respectively, of
$y,y^{\prime}$, and $u_m^{\theta}, v_m^{\theta}$ are solutions to
the equation
\begin{equation}
-u^{\prime\prime}(y) + \left(By+ {2\pi m\over\ell} + \theta
\right)^2 u(y) = zu(y)
\end{equation}
such that $u_m^{\theta}$ is $L^2$ at $-\infty$ and $v_m^{\theta}$
is $L^2$ at $+\infty$; in the denominator we have their Wronskian.
By the argument shift we get
\begin{equation}
u_m^{\theta}(y) = u\left(y+ {2\pi m+\theta\ell\over B\ell} \right)
\nonumber
\end{equation}
and a similar relation for $v_m^{\theta}$, where $u,v$ are the
corresponding  oscillator solutions. Of course, we have
$W(u_m^{\theta}, v_m^{\theta})= W(u,v)$. The functions $u,v$
express in terms of the confluent hypergeometric functions
\cite[Chap.~13]{as}:
\begin{equation}
v(y) = {\it e}^{-By^2/2} U\left({B-z\over 4B}, {1\over 2};By^2
\right)
\end{equation}
away from zero, and $u$ is obtained by analytical continuation in
the $y^2$ variable; together we have
\begin{equation}
\left\lbrace \begin{array}{c} u \\ v \end{array} \right\rbrace(y)
= \sqrt{\pi}\, {\it e}^{-By^2/2} \Bigg\lbrack {M\left({B-z\over
4B}, {1\over 2};By^2 \right) \over \Gamma \left({3B-z\over 4B}
\right)} \pm 2\sqrt{B}y\, {M\left({3B-z\over 4B}, {3\over 2};By^2
\right) \over \Gamma \left({B-z\over 4B} \right)} \Biggr
\rbrack\,.
\end{equation}
From here and Ref.~\cite[Chap.~6]{as} we compute the Wronskian; in
combination with (\ref{Bloch ev}) we get
\begin{eqnarray} \label{G_0}
\lefteqn{G_0(\vec x,\vec x^{\prime};\theta,z) = -\,
{2^{(z/2B)-(3/2)}\over \sqrt{\pi B}\ell}\, \Gamma \left({B-z\over
2B} \right)\, {\rm e}^{i\theta(x-x^{\prime})} } \nonumber \\ &&
\times \sum_{m=-\infty}^{\infty} u\left(y_< + {2\pi m+\theta\ell
\over B\ell} \right) v\left(y_> + {2\pi m+\theta\ell \over B\ell}
\right) \, {\rm e}^{2\pi im(x-x^{\prime})/ \ell}\,.
\end{eqnarray}
As expected the function has singularities which are independent
of $\theta$ and coincide with the Landau levels, i.e.,
$z_n=B(2n\!+\!1), \; n=0,1,2,\dots$. Let us observe first that
each $z_n$ remains to be infinitely degenerate eigenvalue of the
``full" fiber operator $H_{\alpha,\ell}(\theta)$. To this end, one
has to adapt the argument of Refs.~\cite{arb,dmp} to the set of
functions $w^k \sin\left(\pi w\over\ell\right) e^{-B|w|^2/4}, \,
k=0,1,\dots$, with $w:=x+iy$ which vanish at the points of the
array so the conditions (\ref{bc}) are satisfied for them
automatically.

On the other hand, $H_{\alpha,\ell}(\theta)$ has also eigenvalues
away of $z_n$ which we denote as $\epsilon_n(\theta) \equiv
\epsilon_n^{(\alpha,\ell)}(\theta)$. In view of (\ref{Krein}) they
are given by the implicit equation
\begin{equation} \label{implicit}
\alpha = \xi(\vec a_0;\theta,\epsilon)
\end{equation}
and the corresponding eigenfunctions are
\begin{equation} \label{ef}
\psi_n^{(\alpha,\ell)}(\vec x;\theta) = G_0(\vec x,\vec
a_0;\theta,\epsilon_n(\theta))\,.
\end{equation}
In order to evaluate them, we have to assess the convergence of
the series in (\ref{G_0}). Using the asymptotic behavior
\begin{equation}
\left\lbrace \begin{array}{c} u \\ v \end{array} \right\rbrace(y)
= {\it e}^{\mp\{\pm\}By^2/2} \left(\mp\sqrt{B}y \right)^{z-B\over
2B} \left( 1+{\cal O}(|y|^{-2}) \right)
\end{equation}
for $y\to\mp\infty$, we find that the product
$$ s_m := u\left(y_< + {2\pi m+\theta\ell \over B\ell} \right)
v\left(y_> + {2\pi m+\theta\ell \over B\ell} \right) $$
is for $y\ne y^{\prime}$ governed by the exponential term,
\begin{equation}
s_m \sim \exp\left\lbrace {B\over 2} \left(y_<^2-y_>^2\right) +
\left(\theta - {2\pi|m|\over\ell}\right) (y_>-y_<) \right\rbrace
\left(|m|^{-1} + {\cal O}(|m|^{-2})\right)
\end{equation}
as $|m|\to\infty$, while for $y= y^{\prime}$ we have
$$ s_m = -\,{1\over 4\pi}\,|m|^{-1} + {\cal O}(|m|^{-2})\,, $$
so the series (\ref{G_0}) is not absolutely convergent. Summing
now the contributions from $\pm m$ we see that in the limit $x
^{\prime} \to x$ it diverges at the same rate as the Taylor series
of $-(1/2\pi)\ln\zeta$ does for $\zeta\to 0+$. Hence we get
\begin{equation} \label{xi2}
\xi(\vec x;\theta,z) = \sum_{m=-\infty}^{\infty} \left\lbrace
{1-\delta_{m,0}\over 4\pi|m|} -\, {2^{-2\zeta-1} \over \sqrt{\pi
B}\ell}\, \Gamma(2\zeta)\, (uv)\left(y + {2\pi m+\theta\ell \over
B\ell} \right) \right\rbrace\,,
\end{equation}
where $\zeta:= {B-z\over 4B}$. The expression is independent of
$x$, because the regularized resolvent does not change if the
array is shifted in the $x$ direction. We can write it by means of
the first hypergeometric function alone, since
\begin{equation} \label{uv}
(uv)(\xi/\sqrt{B}) = \pi\, {\rm e}^{-\xi^2} \left[{M(\zeta,{1\over
2};\xi^2)^2 \over \Gamma(\zeta+{1\over 2})^2} -4\xi^2
{M(\zeta+{1\over 2},{3\over 2};\xi^2)^2 \over \Gamma(\zeta)^2}
\right]\,,
\end{equation}
where $\xi:= \sqrt{B} \left(y + {2\pi m+\theta\ell \over B\ell}
\right)$.

Spectral bands of our model are given by the ranges of the
functions $\epsilon_n(\cdot)$. Solutions of the condition
(\ref{implicit}) do not cross the Landau levels, because $\xi(\vec
a_0;\theta,\cdot)$ is increasing in the intervals $(-\infty,B)$
and $(B(2n-1),B(2n+1))$ and diverges at the endpoints; this is a
general feature \cite{away}. The spectrum will be continuous away
of $z_n$ if the latter are nowhere constant. In view of the
spectral condition (\ref{implicit}) one has to check that
$\xi(\vec x;\theta,z)$ is nowhere constant as a function of
$\theta$. Notice that each term in (\ref{xi2}) is real-analytic
for real $z$ and the series has a convergent majorant independent
of $\theta$; hence $\xi(\vec x;\cdot,z)$ is real-analytic as well
and one has to check that it is non-constant in the whole
Brillouin zone $[-\pi/\ell,\pi/\ell)$.

Suppose that the opposite is true. Then the Fourier coefficients
\begin{equation}
c_k:= \int_{-\pi/\ell}^{\pi/\ell} \xi(\vec x;\theta,z)\,
e^{ik\ell\theta}\, d\theta
\end{equation}
should vanish for any non-zero integer $k$. Since the summand in
(\ref{xi2}) behaves as ${\cal O}(|m|^{-2})$ as $|m|\to\infty$, we
may interchange the summation and integration. A simple change of
variables then gives
\begin{equation}
c_k = -\, {2^{-2\zeta-1} \over \sqrt{\pi B}\ell}\, \Gamma(2\zeta)
\lim_{M\to\infty} \int_{-\pi(2M+1)}^{\pi(2M+1)} (uv)\left(y +
{\vartheta \over B\ell}\right)\, e^{ik\vartheta}\, d\vartheta\,,
\end{equation}
so
\begin{equation} \label{zero}
\hat{F_y}(k):= \int_{-\infty}^{\infty} F_y(\vartheta)\,
e^{ik\vartheta}\, d\vartheta = 0\,,
\end{equation}
where $F_y(\vartheta):= (uv)\left(y + {\vartheta \over
B\ell}\right)$. The same reasoning applies to any finitely
periodic extension of $\xi(\vec x;\theta,z)$, hence (\ref{zero})
is valid for each non-zero rational $k$. However, the function
decays ${\cal O}(|\vartheta|^{-1})$ and the integral makes sense
only as the principal value. We shall use the above mentioned
asymptotic behavior which implies, in particular,
\begin{equation} \label{expansion2}
F_y(\vartheta) = - {1\over 4\pi \sqrt{1+\vartheta^2}} \,+
f_y(\vartheta)\,, \end{equation}
where $f_y(\vartheta)= {\cal O}(|\vartheta|^{-2})$ uniformly in
$y\in [0,\ell]$. Thus
\begin{equation}
\hat{F_y}(k) = -\,{1\over 2\pi}\, K_0(k) + {\hat f}_y(k)\,,
\end{equation}
see \cite[3.754.2]{gr}. Since $f_y\in L^1$, the second term at the
r.h.s. is continuous w.r.t. $k$ and the same is then true for
$\hat{F}_y$; this means that the relation (\ref{zero}) is valid
for any nonzero $k$. Furthermore, ${\hat f}_y$ is bounded and
$K_0$ diverges logarithmically at $k=0$, hence $\int_{-N}^N
F_y(\vartheta)\, e^{ik\vartheta}\, d\vartheta$ can be bounded by
an integrable function independent of $N$. Then
\begin{equation}
\int_{-\infty}^{\infty} \hat{F}_y(k)\phi(k)\, dk =
\int_{-\infty}^{\infty}\! dk\, \phi(k) \lim_{N\to\infty}
\int_{-N}^N F_y(\vartheta)\, e^{ik\vartheta}\, d\vartheta =
\int_{-\infty}^{\infty} F_y(\vartheta)\, \hat{\phi}(\vartheta)\,
d\vartheta
\end{equation}
holds for any $\phi\in {\cal S}(\mathbb{R})$, i.e., $\hat{F}_y(k)$
is the Fourier transform of $F_y(\vartheta)$ in the sense of
tempered distributions. Since this is a one-to-one correspondence
\cite[Thm.IX.2]{rees}, we arrive at the absurd conclusion that
$F_y=0$. We get thus the following result: \\[2mm]
{\em Theorem.} For any real $\alpha$ the spectrum of
$H_{\alpha,\ell}$ consists of the Landau levels $B(2n\!+\!1), \;
n=0,1,2,\dots$, and absolutely continuous spectral bands situated
between adjacent Landau levels and below $B$. \vspace{2mm}

\noindent Let us remark that during the final stage of the work we
learned about a similar result for a chain of point scatterers in
a three-dimensional space with a homogeneous magnetic field
\cite{agk}. Due to the higher dimensionality, the spectrum is
purely a.c. in that case and has at most finitely many gaps.

The above theorem says a little about the character of the
transport. To get a better idea we solve the spectral condition
(\ref{implicit}) numerically for several values of the parameters.
The results are plotted in Fig.~1 for the second and 21st spectral
band. We see that the bands move downwards with decreasing
$\alpha$ and their profile becomes more complicated with the band
index $n$; a higher $B$ tends to smear the structure.

\begin{figure}
INSERT FIG.~1 (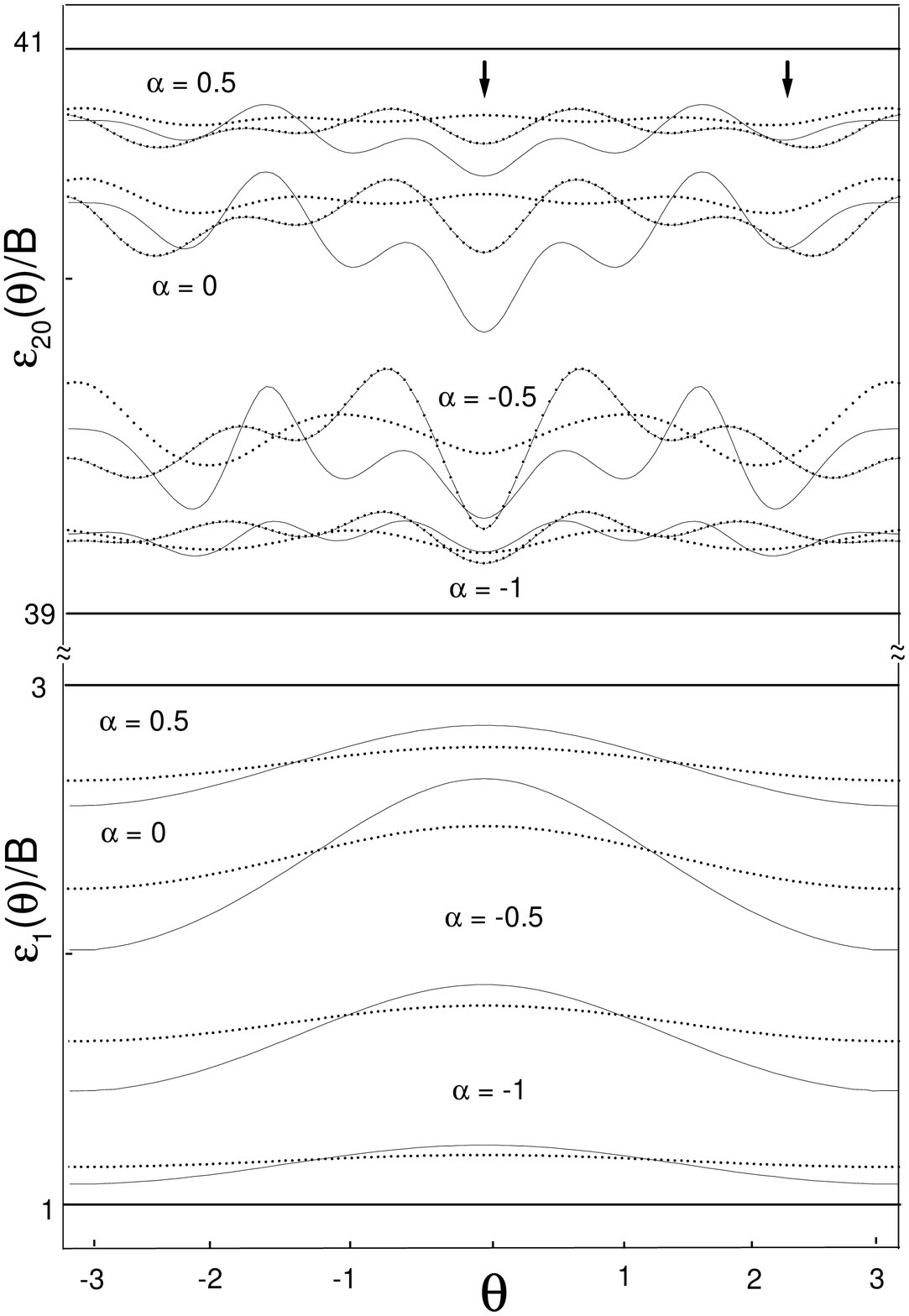)
\caption{The eigenvalues of
$H_{\alpha ,l}(\theta )$ in the units of $B$. (Top) $n=20$, $B=4$
(full line), $B=6$ (full dotted line) and $B=8$ (dotted line).
(Bottom) $n=1$, $B=10$ (full line) and $B=15$ (dotted line). The
thick lines represent the Landau levels.}
\end{figure}

The Bloch functions (\ref{ef}) are in general complex-valued and
yield thus a nontrivial probability current, $\vec \jmath_n(\vec
x;\theta) = 2\,{\rm Im} \left(\bar \psi_n^{(\alpha,\ell)}
(\vec\nabla -i\vec A) \psi_n^{(\alpha,\ell)} \right)(\vec
x;\theta)$. The current pattern changes with $\theta$ oscillating
between a symmetric ``two-way" picture and the situations where
one direction clearly prevails, these extremal behaviors occurring
at the extrema of the corresponding band function. This is
illustrated in Fig.~2. In addition, while the pattern has
predominantly ``laminar" character, in some parts current vortices
may form, mainly in low spectral bands as it is illustrated in
Fig.~3. Similar effects have also been observed in the numerical
analysis of related models mentioned in the introduction.

\begin{figure}
INSERT FIG.~2 (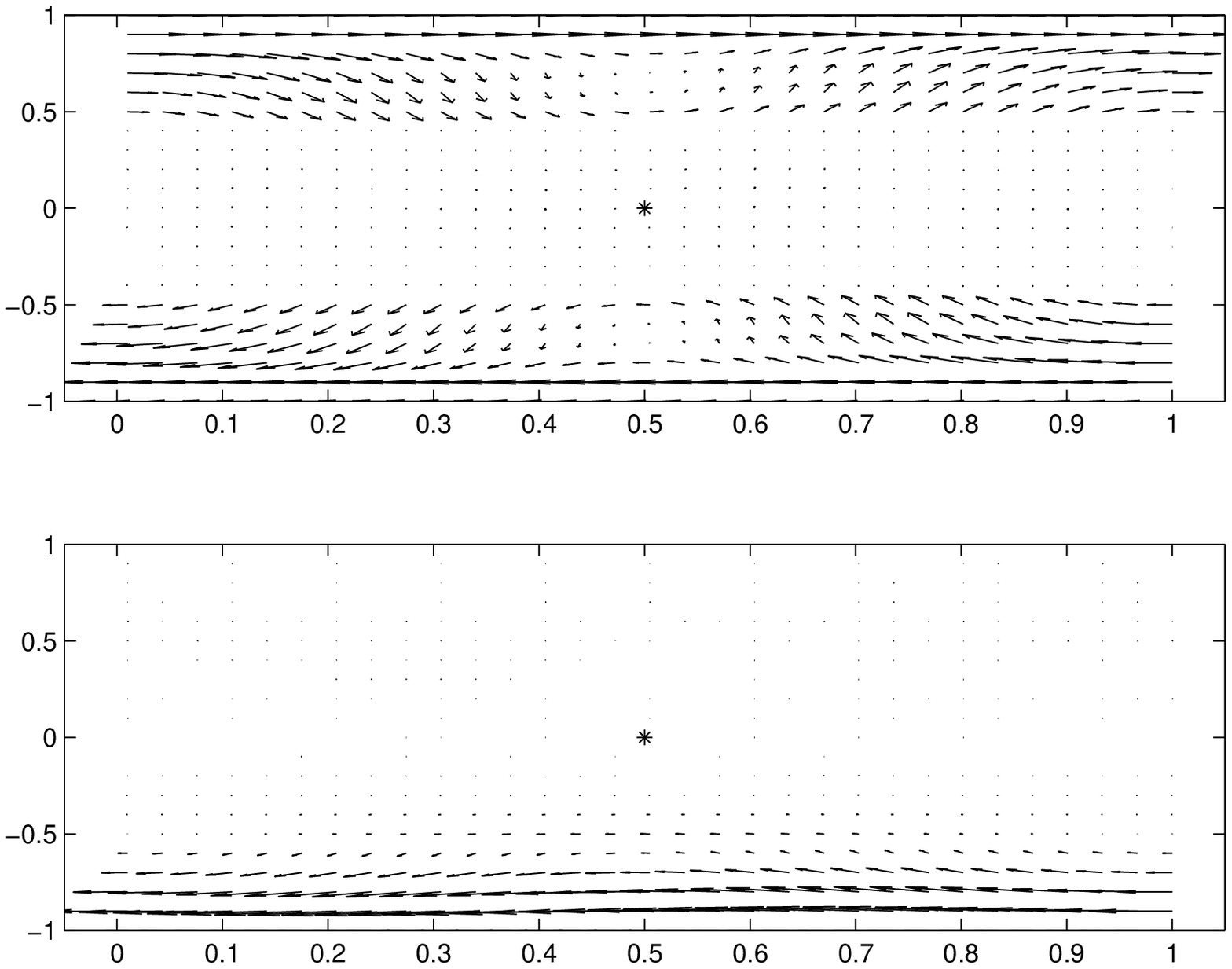)
\caption{Probability current for $n=20$, $B=4$, $\alpha =0.5$ and
two different values of $\theta$ corresponding to the extremal
points of $\epsilon_{20}(\cdot)$: $\theta =0$ (top), $\theta =2.2$
(bottom), see the arrows in Fig.~1. The star  marks the point
perturbation position.}
\end{figure}

\begin{figure}
INSERT FIG.~2 (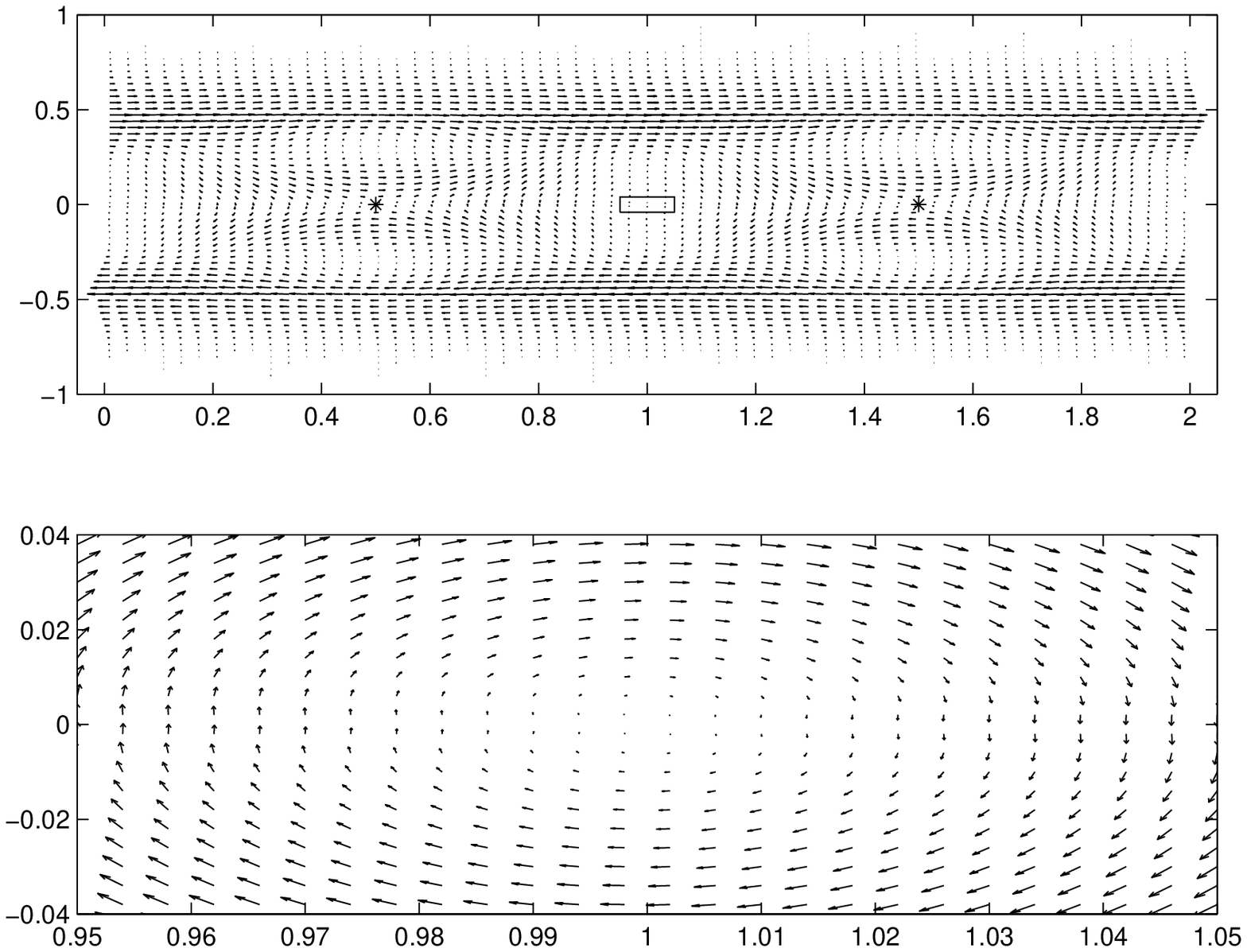)
\caption{Probability current for $n=1$, $B=20$, $\alpha =-1$ and
$\theta =0$. In the bottom graph the inset shows a vortex between
the point perturbations.}
\end{figure}

To sum up the above discussion, we have analyzed the behavior of a
quantum particle in the plane exposed to a homogeneous magnetic
field and interacting with a periodic array of point
perturbations. We have shown that while the Landau levels survive,
the spectrum develops an absolutely continuous part, i.e. a
sequence of spectral bands. Depending on the quasimomentum, the
particle is transported along the array with zero or nonzero mean
longitudinal momentum, and the probability current pattern may
exhibit vortices in some regions.

\subsection*{Acknowledgments}

A.J. wishes to thank his hosts at the Nuclear Physics Institute in
\v{R}e\v{z}, where this work was initiated. The research has been
partially supported by the GAAS grant 1048801. We are grateful to
V.~Geyler for making Ref.~21 available to us prior to publication.

\subsection*{Figure captions}

\begin{description}

\item{\bf Figure 1.} The eigenvalues of
$H_{\alpha ,l}(\theta )$ in the units of $B$. (Top) $n=20$, $B=4$
(full line), $B=6$ (full dotted line) and $B=8$ (dotted line).
(Bottom) $n=1$, $B=10$ (full line) and $B=15$ (dotted line). The
thick lines represent the Landau levels.

\item{\bf Figure 2.} Probability current for $n=20$, $B=4$,
$\alpha =0.5$ and two different values of $\theta$ corresponding
to the extremal points of $\epsilon_{20}(\cdot)$: $\theta =0$
(top), $\theta =2.2$ (bottom), see the arrows in Fig.~1. The star
marks the point perturbation position.

\item{\bf Figure 3.} Probability current for $n=1$, $B=20$,
$\alpha =-1$ and $\theta =0$. In the bottom graph the inset shows
a vortex between the point perturbations.

\end{description}

\end{document}